\documentclass[letterpaper]{article} 
\usepackage{aaai25}  
\usepackage{times}  
\usepackage{helvet}  
\usepackage{courier}  
\usepackage[hyphens]{url}  
\usepackage{graphicx} 
\usepackage{natbib}  
\usepackage{caption} 
\usepackage{algorithm}
\title{Vectorized Attention with Learnable Encoding for Quantum Transformer}
\author {
    Ziqing Guo\textsuperscript{\rm 1,2},
    Ziwen Pan\textsuperscript{\rm 1},
    Alex Khan\textsuperscript{\rm 3},
    Jan Balewski\textsuperscript{\rm 2}
}
\affiliations {
    \textsuperscript{\rm 1}Texas Tech University, TX, USA\\
    \textsuperscript{\rm 2}National Energy Research Scientific Computing Center, Lawrence Berkeley National Laboratory, CA, USA\\
    \textsuperscript{\rm 3}National Quantum Laboratory, MD, USA\\
}
\usepackage{amsmath,amsfonts,amsthm,mathtools,nicefrac,amssymb}

\newcommand{\qcrank}{\textbf{\texttt{QCrank}}}

\newcommand{\eh}{\textbf{\texttt{EHands}}}

\newcommand{\vqdp}{\textbf{\texttt{VQDP}}}
\newcommand{\vnqe}{\textbf{\texttt{VNQE}}}
\usepackage[HTML]{xcolor} 
\definecolor{mycolor}{HTML}{E6521F}
\usepackage{braket}
\usepackage{qcircuit}
\usepackage{environ}
\newcommand{\myqctmp}[2][0.25]{\Qcircuit @C=#2em @R=#1em @!R}
\NewEnviron{myqcircuit}[1][0.25]{\vcenter{\myqctmp[#1]{0.75} {\BODY}}}
\NewEnviron{myqcircuit*}[2]{\vcenter{\myqctmp[#1]{#2} {\BODY}}}

\newcommand{\Ry}{R_y}
\newcommand{\Rz}{R_z}
\newcommand{\CNOT}{\text{CNOT}}

\usepackage{comment}

\usepackage{placeins}

\usepackage{listings} 
\usepackage{xcolor}   

\usepackage{algpseudocode}
\usepackage{booktabs}
\usepackage{tabularx}

\usepackage{subcaption}

\definecolor{codegray}{rgb}{0.5,0.5,0.5}
\definecolor{codepurple}{rgb}{0.58,0,0.82}
\definecolor{backcolour}{rgb}{0.95,0.95,0.92}

\usepackage{adjustbox}

\usepackage{makecell}

\usepackage{caption}

\usepackage{afterpage}

\usepackage[utf8]{inputenc}
\usepackage{pslatex}

\begin{document}
\maketitle

\begin{abstract}
Vectorized quantum block encoding provides a way to embed classical data into Hilbert space, offering a pathway for quantum models, such as Quantum Transformers (QT), that replace classical self-attention with quantum circuit simulations to operate more efficiently. Current QTs rely on deep-parameterized quantum circuits (PQCs), rendering them vulnerable to QPU noise, and thus hindering their practical performance. In this paper, we propose the Vectorized Quantum Transformer (VQT), a model that supports ideal masked-attention matrix computation through quantum approximation simulation and efficient training via vectorized nonlinear quantum encoder, yielding shot-efficient and gradient-free quantum circuit simulation (QCS) and reduced classical sampling overhead.
In addition, we demonstrate an accuracy comparison for IBM and IonQ in quantum circuit simulation and competitive results in benchmarking natural language processing tasks on the state-of-the-art Kingston QPU of IBM. Our noise intermediate-scale quantum (NISQ)-friendly VQT approach unlocks a novel architecture for end-to-end machine learning in quantum computing. 
\end{abstract}
\section{Introduction}
Quantum computing (QC) holds the promise of solving database searching and RSA decrypting problems faster than classical methods, as demonstrated by Grover’s and Shor’s algorithms \cite{grover1996fast, shor1994algorithms}. This advantage stems from the unique quantum mechanical features, such as superposition, entanglement, and interference, which enable rich parallelism and non-classical correlations. Recent advances in quantum data encoding and quantum arithmetic operations \cite{amankwah2022quantum, balewski2024quantum, balewski2025ehands} further extend this capability by efficiently embedding classical data into high-dimensional Hilbert spaces and enhancing representational expressiveness. Vectorized quantum computation dovetails with classical self-attention \cite{vaswani2017attention}, allowing high-capacity models to capture intricate nonlinear dependencies, thereby improving generalization, lowering perplexity, and increasing training efficiency. 

Among the techniques in quantum computing, quantum circuit learning (QCL) \cite{mitarai2018quantum} has emerged as a leading approach for developing data-driven quantum models because of its capacity to PQC for learning complex patterns. By leveraging controlled unitary operations, QCL enables an efficient representation of structured data, facilitating expressive and compact quantum models that support downstream QT models. These benefits are amplified by recent advances in fault-tolerant quantum architectures \cite{acharya2024quantum} and the integration of high-performance computing with QPU \cite{cacheiro2025qmio}, which collectively provide the infrastructure necessary to scale QCL and QT beyond the limitations of the NISQ era.

The core limitation is that the hybrid QCL paradigm currently exploits the restricted usage of unique quantum phenomena, notably the non-cloning theorem and quantum coherence. In practice, PQCs are trained by adjusting the gate rotations to minimize the classical loss computed from repeated projective measurements that collapse the state and discard full-wavefunction information. This is because neural network feedback is drawn only from classical statistics, entanglement and interference are largely unexploited, and intermediate quantum states remain unused. In addition, crosstalk and gate noise further erode coherence, particularly in deep or non-local circuits that exceed NISQ capabilities \cite{huang2024learning}. Although studies of quantum kernel expressivity imply possible benefits \cite{sim2019expressibility}, empirical evidence is inconclusive, and current PQC models typically resemble shallow classical networks and have yet to demonstrate a scalable quantum advantage.

In this study, we introduce a Vectorized Quantum Transformer (VQT) that integrates observable-based quantum arithmetic approximation with nonlinear encoding for the vectorized quantum dot product (VQDP) and a vectorized nonlinear quantum encoder (VNQE). Specifically, our quantum self-attention approach is implemented using uniformly controlled entangling gates with single-qubit rotations and CNOT gates, while an expressive feed-forward network supports nonlinear embedded quantum encoding. The resulting circuit batches key–query pairs, requires no trainable quantum parameters, and remains fully compatible with the current NISQ hardware. Experiments on IBM state-of-the-art superconducting devices demonstrated that multiple quantum heads yield accurate attention scores and efficient model training by leveraging VQDP and VNQE.

\section{Background}
The original idea of the end-to-end QT workflow stems from the proposed AQT \cite{cha2021attention} inspired by the original transformer work \cite{vaswani2017attention} proved by the work demonstrating the 60 qubits Greenberger-Horne-Zeilinger (GHZ) state \cite{carrasquilla2021probabilistic}. This was followed by a fast simulation for an attention mechanism study \cite{gao2023fast} using a Grover search. Such QSAN enables quantum data encoding for natural language processing (NLP) tasks \cite{zheng2023design} using a trainable parameterized quantum circuit (QPC) improved by variational QT using a quantum fourier transform (QFT) kernel \cite{evans2024learning}. 
Such the application of quantum singular value transform (QSVT) \cite{khatri2024quixer} supported quantum signal processing information theory \cite{eldar2002quantum} provides a quantizing dot-product attention mechanism. In addition, the adaptive attention method \cite{chen2025quantum} optimizes the inner dot-product relationship during the PQC training. Although the theoretical protocols here are too finicky, the success of recent NLP applications on quantum computers is likely to make AI algorithms plausible in the future production-scaling quantum computers \cite{widdows2024quantum} with the tool \cite{guo2025q}.
These approaches typically require a learnable long-distance QPC, which is not plausible for near-term noisy quantum computers. 


\subsection{Nonlinear quantum data encoding}
\label{sec:encode}
Basic quantum data encoding techniques can be categorized into basis encoding, angle encoding, and amplitude encoding \cite{weigold2021encoding}. Note that the encoding qubit and repetitive data feeding overhead are listed in Table \ref{tab:encoding}. The encoding techniques stem from the transformation of the mapping of classical data into a quantum state \(x\;\longmapsto\;\ket{\psi(x)}\). 
More generally, let us define a N-dimensional state vector \(x=(x_0,\dots,x_{N-1})^\top\). Here, $x$ as the classical input which maps to the multi-party quantum state \(\frac{1}{\lVert x\rVert_2}\sum_{i=0}^{N-1}x_i\ket{i}\), where \(\lVert x\rVert_2=\Bigl(\sum_{i=0}^{N-1}|x_i|^2\Bigr)^{1/2}\) as the coefficients of each party defined by Born's rule. 
Specifically, given N features, the unitary operation \(U(x)\) acts on the initial fiducial state \(\ket{0}^{\otimes n}\). Here, n is the number of qubits derived by \(n=\bigl\lceil\log_2 N\bigr\rceil\). 
In our model, we inherit the spirit of the expectation-value encoding (EVEN) scheme \cite{balewski2025ehands} because of the natural value constraint correlation between the machine learning (ML) activation function \(\tanh\) and the real value encoding input \(x_i\in [-1,1]\) (see details in Section \ref{sec:VQDP}). The EVEN encoding fits the category of angle encoding because the constructed quantum state can be subjected to single qubit \(\Ry(\theta_i)\) rotation, where \(\theta_i = \arccos(x_i)\). 
\begin{table}[htbp]
\centering
\caption{Resource scaling for three common data encoding strategies when processing $M$ samples with $N$ features each.}
{
\begin{tabular}{lcc}
\toprule
\multicolumn{1}{c}{\bf Encoding} & \bf Number of Qubits & \bf Number of Passes \\
\midrule
Basic       & $N$               & $1$ \\
Amplitude   & $\log_{2}(MN)$    & $1$ \\
Angle       & $N$               & $M$ \\
\bottomrule
\end{tabular}}
\label{tab:encoding}
\end{table}

\subsection{Quantum State Evolution}
To build the quantum circuit \cite{nielsen2010quantum} model for computation, in which computation is a sequence of quantum gates including single-qubit operation, two-qubit entanglement gate, and multi-qubit gates. In general, we use unitary evolution to preserve the probability of the quantum state, namely Hamiltonian $\mathcal{H}$ \(|\psi(t)\rangle = e^{-iHt/\hbar} |\psi(0)\rangle\). Specifically, the quantum state evolution can be denoted by the observable \(O = \langle \psi|H(t)|\psi\rangle\). Here, $\mathcal{H}$ is a sequence of Hermitian matrices.
Note that \(\Ry\) gate specifically introduce a real, continuous rotation of the qubit about y-axis of the Bloch sphere by the angle \(\theta\)
\begin{equation}
\Ry(\theta)=
\bigl[\!
  \begin{matrix}
    \cos(\tfrac{\theta}{2}) & -\sin(\tfrac{\theta}{2})\\
    \sin(\tfrac{\theta}{2}) &  \cos(\tfrac{\theta}{2})
  \end{matrix}
\!\bigr],
\qquad
\Ry\!\bigl(\tfrac{\pi}{2}\bigr)=
\bigl[\!
  \begin{smallmatrix}
    \cos(\tfrac{\pi}{4}) & -\sin(\tfrac{\pi}{4})\\
    \sin(\tfrac{\pi}{4}) &  \cos(\tfrac{\pi}{4})
  \end{smallmatrix}
\!\bigr].
\end{equation}
By following the EVEN encoding, we consider two classical input values \( \{x_0,x_1\} \) in the range \([-1, 1]\). The encoding process involves feeding each value \( x_i \) to the parameterized rotation gates \( \Ry(\theta_i) \), which are then applied to a 2-qubit quantum state initialized in \( \ket{00} \). Here, we denote the quantum state as \(\ket{\phi_i}\) input composed of tensor products of unitary operations acting on the initial state.
\begin{align}
\ket{\phi_i}
  = \ket{00} \bigl[ \Ry{\theta_0}\otimes\Ry{\theta_1} \bigr] 
  = \frac12
  \begin{bmatrix}
      \sqrt{1+x_0}\,\sqrt{1+x_1} \\[4pt]
      \sqrt{1+x_0}\,\sqrt{1-x_1} \\[4pt]
      \sqrt{1-x_0}\,\sqrt{1+x_1} \\[4pt]
      \sqrt{1-x_0}\,\sqrt{1-x_1}
  \end{bmatrix},
\label{eq:x2theta}
\end{align}
thereby achieving classical information embedded into the two-qubit quantum state. The encoding scheme can be represented as a quantum circuit.
\begin{equation}
  \centering
      \begin{myqcircuit}
        & \barrier{1} & \qw & \gate{\Ry{\theta_0}} & \qw \\
        &            & \qw & \gate{\Ry{\theta_1}} & \qw
        \inputgroup{1}{2}{.75em}{\ket{0^i}}
      \end{myqcircuit}
      \hspace{.2em}
      \Bigg\}\ket{\phi_e},
      \quad i\in\{0,1\},\;
      \theta_i=\cos(x_i),
  \label{eq:Ry_encoding}
\end{equation}
Furthermore, the entanglement gate enables state modification between multiple qubits. For instance, one of the best techniques to showcase the state changes is to add a \(\CNOT\) gate after the encoded state \(\ket{\phi}\) with an \(Rz\) gate. By doing this, the second qubit acquires a phase that is conditioned on the logical value of the first qubit. Consequently, the two qubits are no longer in a simple product state. Here, the \(\ket\phi\) state from \eqref{eq:x2theta} can be evolved as
\begin{equation}
\begin{aligned}
    \ket{\phi_\Pi} &:= \ket\phi_e\cdot\text{CNOT}_{0,1}\cdot \left[ I \otimes  \Rz{\nicefrac{\pi}{2}} \right] = \frac{e^{-i\frac{\pi}{4}}}{2} \left[\begin{smallmatrix}
           \sqrt{1 + x_0} \sqrt{1 + x_1} \\
        i\,\sqrt{1 + x_0} \sqrt{1 - x_1} \\
        i\,\sqrt{1 - x_0} \sqrt{1 - x_1} \\
           \sqrt{1 - x_0} \sqrt{1 + x_1} \\
    \end{smallmatrix}\right]\\
    &=e^{-i\pi/4}\,
\bigl(
      \cos\tfrac{\theta_0}{2}\cos\tfrac{\theta_1}{2}\,\ket{00}
    + i\cos\tfrac{\theta_0}{2}\sin\tfrac{\theta_1}{2}\,\ket{01}
\\
    &+ i\sin\tfrac{\theta_0}{2}\sin\tfrac{\theta_1}{2}\,\ket{10}
    +   \sin\tfrac{\theta_0}{2}\cos\tfrac{\theta_1}{2}\,\ket{11}
\bigr).
    \end{aligned}
    \label{eq:eh}
\end{equation}

To expand the encoding into a broader data permutation scenario, the quantum circuit can be represented by a sequence of multiple controlled qubit gates, such as the controlled rotation gate. Because the Pauli gates group follows the thoery of commuting with itself based on Baker–Campbell–Hausdorff formula. 
A motivation example is to represent the classical information by using the permutated controlled Y rotation gate \((CRY)\)
\begin{equation}
\quad
\begin{myqcircuit}
\lstick{x_1} & \ctrlo{1} & \ctrl{1} & \qw \\
\lstick{x_2}   & \gate{R_y(\theta_1)} & \gate{R_y(\theta_2)} & \qw 
\end{myqcircuit}\rightarrow 
\begin{aligned}
x_1 &= \theta_1 + \theta_2 \\
x_2 &= \theta_1 - \theta_2 \\
\end{aligned}.
\label{eq:cry}
\end{equation}

\subsection{Quantum Measurement}
\label{sec:meas}
In the decoding phase, quantum measurement provides a definitive classical state that collapses from an uncertain quantum state. A common strategy is to perform a Z-basis measurement on the quantum state \cite{camps2022quantum}, which leads to a projection onto the computational basis \(\ket{0}, \ket{1}\). 
The corresponding \emph{decoding} quantum circuits are depicted in (b) and (c) of Circuit 1 \cite{balewski2025ehands}, where 
$\raisebox{0.2em}{\Qcircuit @C=0.4em {& \qw & \measuretab{Z}}} $
denotes measuring the EV of the corresponding qubit in the $Z$-basis.

The expectation value is constrained by the shot error, namely, the repetitive execution of each quantum circuit. Note that the quantum measurement error can be largely reduced by sufficient Monte Carlo sampling simulations. Specifically, \(O\) is a Hermitian single–qubit observable with spectral norm \(\|O\|_{\infty}\le 1\), and \(\mu=\operatorname{Tr}(O\rho)\) is its expectation in the state \(\rho\).
A single projective measurement returns a bounded random variable \(X\in[-1,1]\); after \(M\) i.i.d.\ shots, the estimator 
\(
\bar X_M=\frac{1}{M}\sum_{k=1}^{M}X_k
\)
satisfies Hoeffding’s inequality,
\(
\Pr\left[|\bar X_M-\mu|\ge\varepsilon\right]\le 
2e^{-2M\varepsilon^{2}}.
\)


\section{Method}
\begin{figure}[t]
    \centering
    \includegraphics[width=0.99\linewidth]{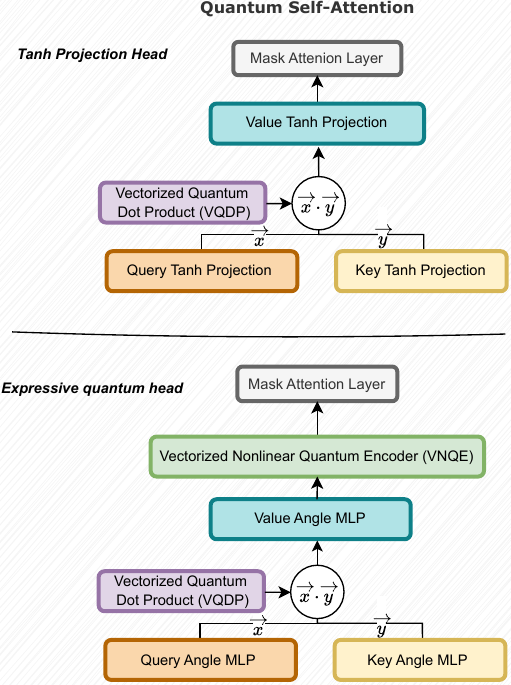}
    \caption{Quantum multi-head attention comprises two components: the tanh projection head employs \(\tanh\) projection matrices to map the input into a quantum-compatible representation, whereas the expressive quantum head provides the learnable angle-encoding multi-layer perceptron (MLP) that encodes the data quantum-mechanically.}
    \label{fig:heads}
\end{figure}

We refer to Fig. 5 in \cite{smaldone2025hybrid} for an overview of the architecture of the quantum transformer learning model. 
Following the spirit of the hybrid quantum transformer, we encode the classical input with positional and token encoders formulated by adding together to construct the input embedding for the feedforward network with layer normalization and softmax for output probability. However, the multi-head attention layer is neither classically nor quantumly unique. 
The classical attention score is calculated using a sequence of tokens. Note that, input token embeddings \(\{t_1, t_2, \cdots, t_i\}\) are augmented with the positional encoding \(\{p_1, p_2,\cdots, p_i\}\) to form composite embeddings
\begin{equation}
   \mathrm{}{z}_i = \mathrm{t}_i + \mathrm{p}_i.
\end{equation}
The remainder of the method section follows our proposed quantum self attention method, tanh projection head, and expressive quantum head, as indicated in Figure \ref{fig:heads}. 

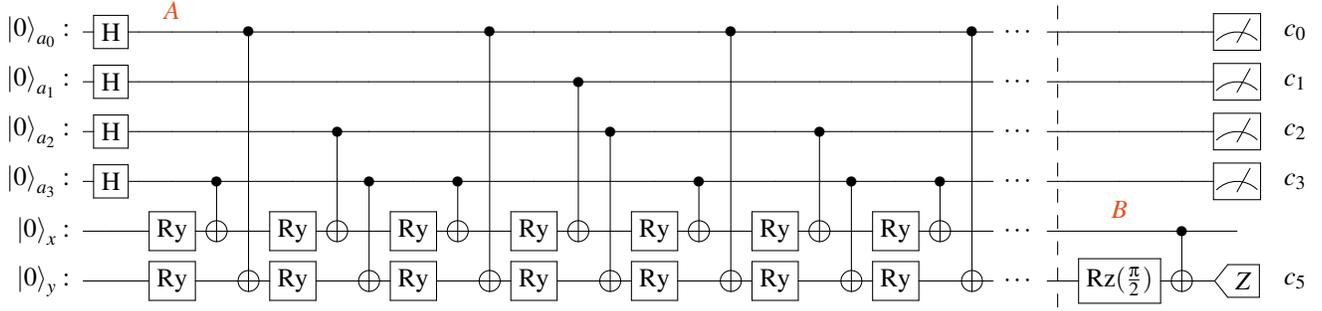
\begin{figure*}[t]
\hspace{10mm}
\scalebox{0.99}{
\Qcircuit @C=0.4em @R=0.2em @!R { \\
	 	 & \lstick{\ket{0}_{a_0} :  } & \gate{\mathrm{H}}  & \qw &\ustick{\textcolor{mycolor}{A}} \qw & \qw & \ctrl{5} & \qw & \qw & \qw & \qw & \qw & \ctrl{5} & \qw & \qw & \qw & \qw & \qw & \ctrl{5} & \qw & \qw & \qw & \qw & \qw & \ctrl{5}  & \qw & \push{\cdots~~}  \barrier[0em]{5} & \qw & \qw  & \qw & \qw & \qw & \meter &\rstick{c_0}  \\
	 	 & \lstick{\ket{0}_{a_1} :  } & \gate{\mathrm{H}} & \qw & \qw & \qw & \qw & \qw & \qw & \qw & \qw & \qw & \qw & \qw & \ctrl{3} & \qw & \qw & \qw & \qw & \qw & \qw & \qw & \qw & \qw & \qw & \qw&\push{\cdots~~}  & \qw    & \qw  & \qw& \qw& \qw  & \meter &\rstick{c_1}  \\
	 	 & \lstick{\ket{0}_{a_2} :  } & \gate{\mathrm{H}} & \qw & \qw & \qw & \qw & \qw & \ctrl{2} & \qw & \qw & \qw & \qw & \qw & \qw & \ctrl{3} & \qw & \qw & \qw & \qw & \ctrl{2} & \qw & \qw & \qw & \qw  & \qw &\push{\cdots~~} & \qw & \qw  & \qw& \qw & \qw & \meter &\rstick{c_2}\\
	 	 & \lstick{\ket{0}_{a_3} :  } & \gate{\mathrm{H}} & \qw & \qw & \ctrl{1} & \qw & \qw & \qw & \ctrl{2} & \qw & \ctrl{1} & \qw & \qw & \qw & \qw & \qw & \ctrl{1} & \qw & \qw & \qw & \ctrl{2} & \qw & \ctrl{1} & \qw &   \qw &\push{\cdots~~} & \qw & \qw & \qw  & \qw & \qw & \meter &\rstick{c_3}\\
	 	 & \lstick{\ket{0}_{x} :  } & \qw & \qw & \gate{\mathrm{Ry}} & \targ & \qw & \gate{\mathrm{Ry}} & \targ & \qw & \gate{\mathrm{Ry}} & \targ & \qw & \gate{\mathrm{Ry}} & \targ & \qw & \gate{\mathrm{Ry}} & \targ & \qw & \gate{\mathrm{Ry}} & \targ & \qw & \gate{\mathrm{Ry}} & \targ & \qw & \qw & \push{\cdots~~} & \qw  & \qw & \ustick{\textcolor{mycolor}{B}} \qw &\ctrl{1}& \qw& \qw \\
	 	 & \lstick{\ket{0}_{y} :  } & \qw & \qw & \gate{\mathrm{Ry}} & \qw & \targ & \gate{\mathrm{Ry}} & \qw & \targ & \gate{\mathrm{Ry}} & \qw & \targ & \gate{\mathrm{Ry}} & \qw & \targ & \gate{\mathrm{Ry}} & \qw & \targ & \gate{\mathrm{Ry}} & \qw & \targ & \gate{\mathrm{Ry}} & \qw & \targ & \qw  & \push{\cdots~~} & \qw  & \qw  &\gate{\mathrm{Rz(\frac{\pi}{2})}}  &\targ & \qw  & \measuretab{Z} &\rstick{c_5} 
\\ }}
\caption{ Example quantum circuit computing vectorized product (\vqdp) $x_i \cdot y_i$  for the batch size of 32 $(x_i, y_i)$ pairs. It uses four qubits for addresses and two qubits for input values of $x_i,y_i$. \qcrank\ (A) encoding is before barrier and \eh\ (B) multiplier after it. The expectation value of register $c_4$ yields $x_i \cdot y_i$ for index $i$ given by classical registers $c_0,...,c_3$. Here $a_0,...a_3$ are address qubits ($nq_{addr}$) and $x,y$ are data qubits ($nq_{data}$).
}
\label{fig:circ_QC+EH}
\end{figure*}

\subsection{Quantum Attention Score}
\label{sec:VQDP}
The quantum universal polynomial approximation (\eh) and quantum data encoding (\qcrank) were introduced in \cite{balewski2024quantum,balewski2025ehands}. The essential quantum circuit for calculating the element-wise matrix dot product is arithmetic multiplication, as shown in Figure. \ref{fig:circ_QC+EH}. 
The proposed \vqdp\ protocol receives the (non-transposed) query and key tensors \(Q,K\!\in\!\mathbb{R}^{B\times T\times d}\), where B is batch size, T is the sequence length, and d is the feature size. To evaluate the \(N=B\,T^{2}\) inner products \(Q_{b,i}\!\cdot\!K_{b,j}\) it prepares \(n_{\mathrm{addr}}=\lceil\log_{2}N\rceil\) address qubits in a uniform superposition together with two data qubits.  \qcrank\ A enables, for every address \(\ell\), the real pair \(\bigl(Q_{b,i,k},\,K_{b,j,k}\bigr)\) (\(\ell\!\leftrightarrow\!(b,i,j)\)) onto the data qubits. \eh\ B then converts the expectation value \(\langle Z\rangle\) of the second data qubit into the product \(x_{\ell}y_{\ell}\) (see Eq. \eqref{eq:eh}).  Executing this shallow circuit once per feature \(k\in\{1,\dots,d\}\) and averaging over \(S\) shots yields \(P_{\ell k}\!\approx\!x_{\ell}y_{\ell}\); classically summing over \(k\) and reshaping \(\ell\) back to \((b,i,j)\) produces the attention matrix \(A\in\mathbb{R}^{B\times T\times T}\).  Hence, the approach performs the full \(QK^{\top}\) computation with \(n_{\mathrm{addr}}\!+\!2\) qubits, incurring the overhead transformation from the classical \(O(Nd)\) multiply–accumulate floating point operations per second (FLOPs) cost to circuit layer operations per second (CLOPs) cost at \(O\!\left(\log N\right)\).
We denote our method is classically equalized to matrix multiplcation method provided by pytorch 
when the quantum Monte Carlo shots suffice for attaining plausible results. This is nontrivial, as proved in Appendix A. Note that overflow addresses introduced by the power-of-two padding are loaded with all-zero vectors; therefore, any state beyond the \(B T^{2}\) valid pairs contributes nothing to the computed dot products.
We recall Eq. \eqref{eq:x2theta}, the second qubit final measured result $\ket{\phi_{output}}$ is measured by $\mathcal{O}_z$ (Z-basis measurement).
\begin{equation}
O_{z} := I \otimes \sigma_z = \begin{bmatrix}1 & 0 & 0 & 0 \\
        0 & -1 & 0 & 0 \\
        0 & 0 & 1 & 0 \\
        0 & 0 & 0 & -1 \\\end{bmatrix}   
\label{eq:Oz}
\end{equation}
Hence, according to Eq. \eqref{eq:Oz} and Eq. \eqref{eq:x2theta}, we proceed the final observable result
\begin{equation}
    \langle \mathcal{O}\rangle = \bra{\psi_{\Pi}} \mathcal{O}_{z} \ket{\psi_\Pi}.
    \label{eq:phio}
\end{equation}
Note that we only measure the second data qubit because the $\CNOT$ gate does not affect the first qubit. 
Given Eq. \eqref{eq:Oz} and Eq. \eqref{eq:phio}, the output state analytical result is $\ket{\phi_{o}} = \overrightarrow x_i\cdot \overrightarrow y_i$ which is the simplification of $\langle{\phi_o}\rangle$. But in quantum circuit simulation (QCS), the numerical expectation output using Z basis which can be typically reconstructed by the quantum Monte Carlo shots denoted by 
\begin{equation}
    \langle\mathcal{Z}\rangle = \frac{n_0 - n_1}{n_0 + n_1}.
\end{equation}
We would like to emphasize $n_0$ and $n_1$ are the number of shots corresponding to the $\ket{0}, \ket{1}$ measurements, respectively, which vary based on the real quantum hardware for each run. 

\subsection{Tanh Projection Head}
\label{sec:tanh}
For the classical input to the quantum attention head, we chose tanh to narrow the input down to output within -1 to 1 because we utilize \texttt{$\arccos$} to encode the data into angles for quantum circuits. The benefits become clearer when the model uses a vectorized quantum dot product to calculate the attention score because of the polarized distributed multiplication. Additionally, recall that it is possible to replace the tanh projection into the QAF \cite{act3} but in the case of nonlinear quantum encoding (see Section \ref{sec:encode}), the entire sequence of the data serves as the negative value contributing to the probability of the classical output after absolute probability calculation using Born's rule, where the QAF has suboptimal behavior. 
Simply, by constructing a tanh projection head, one can typically confront the $\tanh$ function encoding the query and key matrices through the \vqdp. 
The remaining architecture follows with a value and self-attention matrix connected with another \vqdp\ layer to compare the similarity between the attention and value matrices. Therefore, the entire quantum overhead is twice the number of attention matrices. 

\subsection{Expressive Quantum Head}
\label{med:eqh}
\begin{figure}
    \centering
    \includegraphics[width=0.9\linewidth]{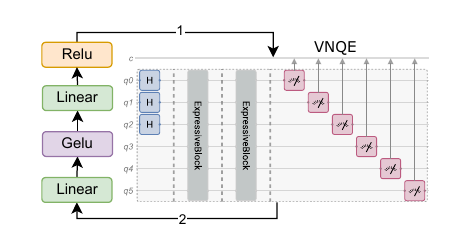}
    \caption{Default three-by-three expressive quantum head. The left deep learning network is the simplified representation of AngleMLP and right quantum circuit is vectorized nonlinear quantum encoder (expressive block shown in Figure 1 of \qcrank).}
    \label{fig:EQH}
\end{figure}
For the expressive layer, we first model a simple multi-layer perceptron named AngleMLP, as shown in Figure \ref{fig:heads} detailed in Figure \ref{fig:EQH}. Arrow $\mathrm{1}$ represents the untrained sequence of three-dimensional tensors, which are parameterized by the batch size and sequence length, as output from AngleMLP and encoded into \vnqe. Arrow $\mathrm{2}$ indicates an approximate observable, depicted as a superpositioned data output, which serves as the input for training AngleMLP. The quantum circuit is composed of the address and data qubits \(nq_{addr}, nq_{data}\) representing the number in which the matrix corresponding to the column index is followed by the number of expressive blocks consisting of permutated controlled $\Ry$ gates, where the gate enables complex phase transition that is encoded by the input data angles (notation details referred in Eq. \eqref{eq:cry}). Consequently, the tensor can be compressed into a vectorized quantum circuit with one pass nonlinearly, which saves the classical nonlinear conversion time from $O(n)$ passes to $O(1)$. Given the unfreezed AngleMLP, during the training phase, the MLP layer enables parameter adjustment before the quantum data encoding, thereby promising zero-gradient transformation because the angles of each individual gate are exactly mapped with the tuned MLP output. This leads to QT improvement, in which quantum computing fulfills the nonlinearity encoding combined with classical neural networks trained using backpropagation.


For \vnqe, we use qubits to encode the quantum embedded hidden dimension that is calculated with 
\begin{equation}
Q_{dim} = 2^{nq_{addr}}*nq_{data}.
\label{eq:qdim}
\end{equation}
To maximize the encoded dimension, Eq. \eqref{eq:qdim} can be denoted by
\begin{equation}
 Q_{\text{dim}}^{\text{max}} = 2^{N-1},     
\end{equation}
where $nq_{\text{addr}} = N-1$, $nq_{\text{data}} = 1$, and N is the total number of qubits. In the ideal scenario, for a 156 IBM Marrakesh QPU, the encoded quantum dimension is $4.56719262 \times 10^{46}$; however, we save this for future research because the long-distance qubit crosstalk error cannot be eliminated in today's QPUs. Meanwhile, we also provide the lower bound for qubits concerning the encoded quantum dimension 
\begin{equation}
    N_{\min}(Q_{\text{dim}}) = \min_{\substack{1 \leq k \leq \log_2 Q_{\text{dim}} }} \left[\, k + \frac{Q_{\text{dim}}}{2^k} \,\right],
\end{equation}
where $Q_{\text{dim}} - 2^{nq_{\text{addr}}} \geq 1$ and is integer.
The loose lower bound is defined as 
\begin{equation}
    N_{\min} = \lceil \log_2(Q_{\text{dim}}) \rceil + 1 
\end{equation} because we assume the encoded quantum dimension is slightly larger than the required dimension. In our simplest scenario, requiring an encode quantum dimension of 32, we set the default number of qubits to 6.

\subsection{Metrics}

Owing to the scarcity of peer-reviewed research articles, there is currently no standardized model evaluation for QT models. However, drawing from classical evaluation methods, we propose the concept of quantum perplexity (QPL) to assess the complexity of quantum models and evaluate the quality of QT prediction. Given by the perplexity definition $PP(p) = b^{H(p)}$, H is the entropy of the distribution, the QPL metric is defined as follows, given a sequence of tokens \( x_1, x_2, ..., x_N \)
\begin{equation}
    \text{QPL} = \exp\left(-\frac{1}{N} \sum_{t=1}^N \log p(x_t | x_{<t})\right).
    \label{eq:ppl}
\end{equation}
Note that $p_{(x_t | x_{<t})}$ is informed by QCS because the final prediction is based on the measured bit string outcomes for reconstruction.

\section{Experiment}
\label{sec:exp}
We show that VQT is a NISQ-friendly quantum model that enables an end-to-end quantum gradient-free workflow on noisy quantum hardware. By leveraging Perlmutter’s state-of-the-art GPU node, VQT supports classical optimization fully accelerated by running on GPU nodes at the scale \cite{nersc2025perlmutter_arch}. Below, we demonstrate the model performance for two representative cases: (1) quantum attention analysis and (2) model performance evaluation.


\subsection{Quantum Attention Analysis}
\paragraph{Multiplication statistical behaviour}
In \vqdp\ ideally one obtains 
\(\langle Z\rangle_{\mathrm{ideal}} = \overrightarrow x\,\cdot \overrightarrow y\).
\begin{figure}[htb]
\centering
\includegraphics[width=0.99\columnwidth]{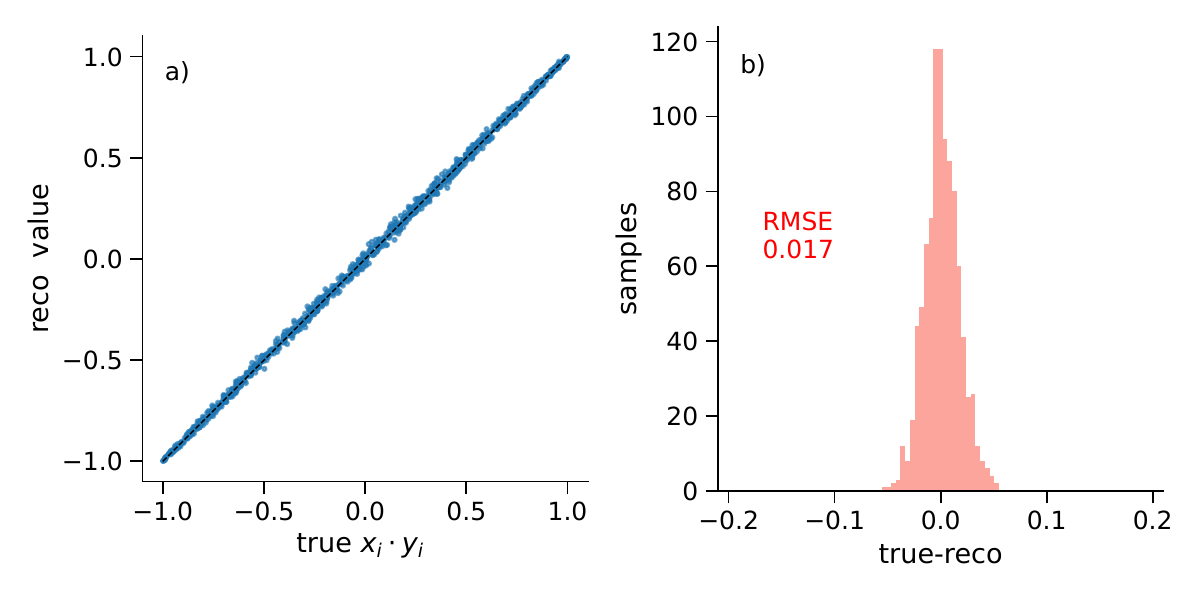}
\caption{
 Accuracy of computation $x_i \cdot y_i$  using 6 qubits circuit shown in Fig. ~\ref{fig:circ_QC+EH} from  the ideal Qiskit simulator with $N_{shot}$=80,000. a) Correlation between true and measured values, averaged over 30 batches of 32  randomly sampled $(x_i, y_i)$ pairs. b) distribution residuals have std dev  0.017,   which scales with $\nicefrac{1}{\sqrt{N_{shot}}}$.
}
\label{fig:resIdeal}
\end{figure}
\begin{figure}[htb]
\centering
\includegraphics[width=0.99\columnwidth]{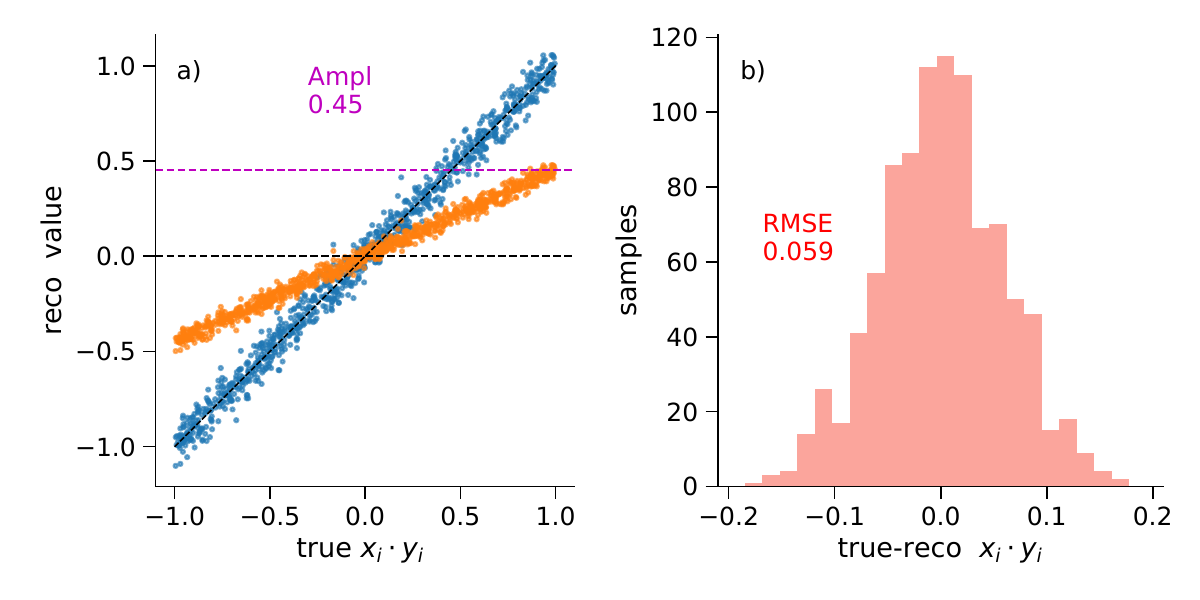}
\caption{ Results of computation of product  $x_i \cdot y_i$  on IBM  Kingston using the same configuration as in Fig~\ref{fig:resIdeal}. a) Correlation between the truth and raw measured data is shown in orange. The blue data show the correlation after scaling of the measurement by a factor of 2.22. b) The residuals for the scaled measurement have std dev  0.059. 
}
\label{fig:resHW}
\end{figure}

\begin{table*}[ht]
\centering
\begin{tabular}{@{}ccc|ccc|ccc|c@{}}
\toprule
 \textbf{batch }& \textbf{num }&  \textbf{num } & \multicolumn{3}{c}{\textbf{Ideal}}& \multicolumn{3}{c}{\textbf{IBM Kingston}} & {\textbf{IonQ Aria-1}$^\dag$} \\
\cmidrule(lr){4-6} \cmidrule(lr){7-9} \cmidrule(lr){10-10}
\textbf{ size} & \textbf{ qubits} & \textbf{ shots} 
 & \textbf{num CX$^\star$} & \textbf{CX depth} & \textbf{RMSE} 
 & \textbf{num CZ} & \textbf{CZ depth} & \textbf{RMSE}  & \textbf{RMSE}\\
\midrule
4   & 4 & 10K   & 9   & 5   & 0.017 &  18  & 14  & 0.022 & 0.057 \\
8   & 5 & 20K  & 17  & 9   & 0.016  &  29  & 25  & 0.023 & 0.155 \\
16  & 6 & 40K   & 33  & 17  & 0.017 &  78  & 69  & 0.037 & 0.215\\
32  & 7 & 80K   & 65  & 33  & 0.017 &  164 & 121 & 0.059 & -- \\
64  & 8 & 160K  & 129 & 65  & 0.016 &  348 & 249 & 0.274 & -- \\
128 & 9 & 320K  & 257 & 129 & 0.016 &  704 & 514 & -- & --   \\
\bottomrule
\end{tabular}
\caption{The characterization of circuits employed in simulations and experiments, along with the achieved accuracy for IBM and IonQ quantum platforms, is presented. Note that CX$^\star$ indicates that the number of CX (CNOT) gates for the corresponding ideal simulators and transpiled IonQ Aria-1 are configured identically. IonQ$^\dag$ has all-to-all connectivity with sequential gate implementation, rendering the CX gate count the primary determinant of execution efficiency, whereas IBM’s parallel architecture makes the native controlled-Z (CZ) gate depth more salient. The root mean squared error (RMSE) serves as a confidence interval for real hardware. Results denoted by $-$ are considered to be meaningless.}
\label{tab:res}
\end{table*}
We observed that the noisy QPU is also able to produce low-error multiplication results with amplitude correction, as shown in Figure \ref{fig:resHW} compared with Figure \ref{fig:resIdeal}. 
Note that halving the error requires approximately four times as many shots.
The experimental settings and results with 30 iterated batches are presented in Table \ref{tab:res}, where IBM outperforms IonQ's hardware in terms of RMSE.

\paragraph{Quantum\,vs.\,classical attention}

\begin{figure}[h]
    \centering
    \includegraphics[width=0.99\linewidth]{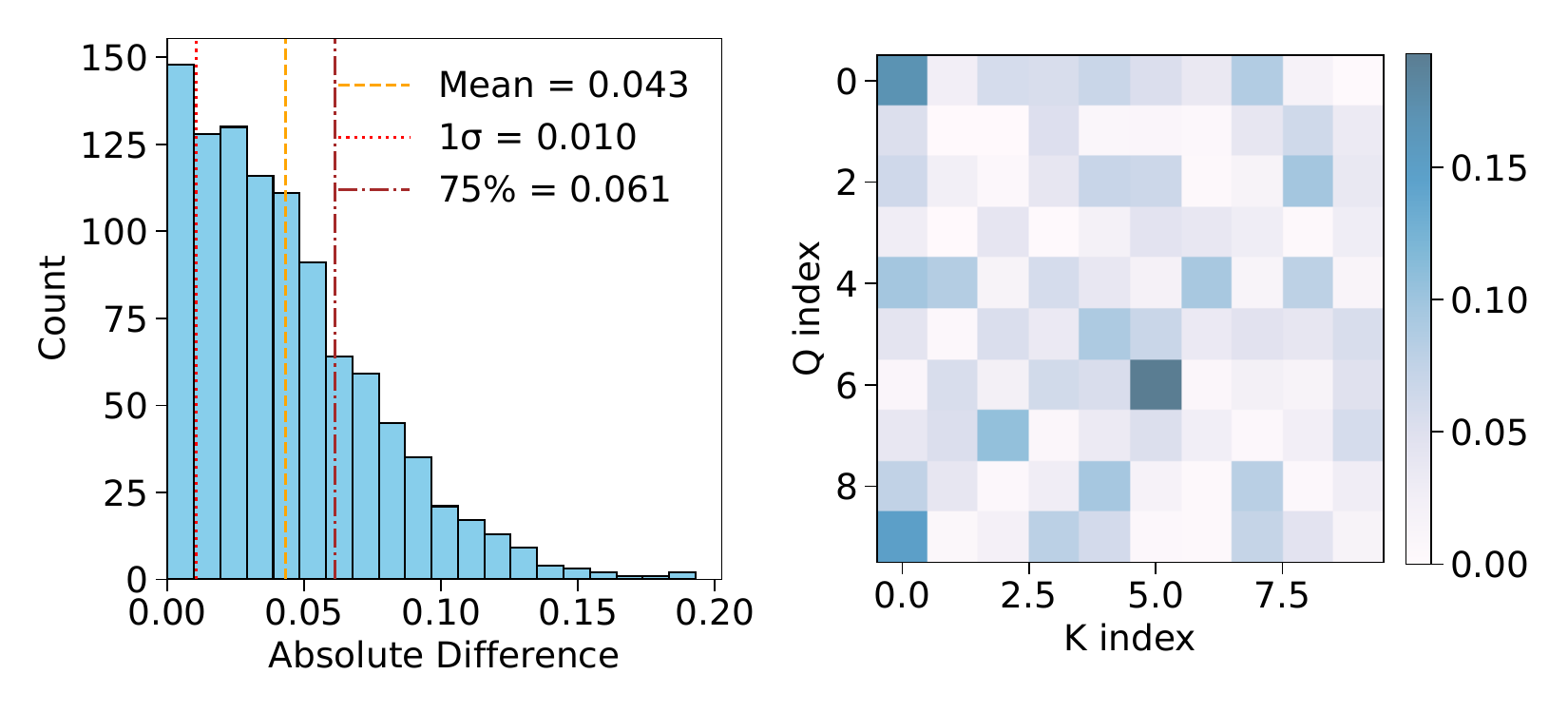}
    \caption{The left panel depicts the distribution of absolute deviations between quantum and classical attention scores computed over all query–key pairs and batches, while the right panel visualizes the corresponding error matrix, in which lighter shades denote smaller discrepancies.}
    \label{fig:cc_vs_qc}
\end{figure}

We show that, with an input size of (10,10,10) corresponding to batches, sequence length, and features, the error between classical attention and \vqdp\ is constantly below 1.2\% with 3.0 million shots. Here, although the average two-qubit gate error for the QPU is approximately $2.5e^{-3}$ (note that the real-time error varies; see 
, our result indicates that the VQDP method allows for competitive results compared with the classical counterpart.

In addition, replacing the classical attention kernel with a quantum sampler yields conceptual benefits. First, the variance in Eq. \eqref{eq:variance} acts as a data-dependent stochastic regularizer which stabilizes training in regimes where the classical model tends to overfit (see Section \ref{sec: me}). Second, the quantum mechanism provides nonlinear kernels whose functional forms can be modified in hardware (choice of measurement basis or additional controlled phases) without changing the classical model architecture.
Additionally, Table \ref{tab:analysis} compares the scaling rule of the classical matrix-multiplication (MatMul) algorithm with \vqdp.
\begin{table}[htbp]
\centering\small
\caption{Complexity analysis for classical and quantum attention
(\(s\)=shots, \(g\)=circuit depth, \(t_{1\mathrm Q}\)=one-qubit gate time, \(t_{\mathrm{CX}}\)=CNOT time, \(t_{\mathrm{ro}}\)=read-out time).  
The wall-clock cost of VQDP is ultimately limited by the circuit-layer-operations-per-second (CLOPS) rate of the QPU \cite{clops}.}
\label{tab:analysis}
\begin{tabular}{@{}lcc@{}}
\toprule
 & \textbf{Classical matmul} & \vqdp\ \\
\midrule
Time        & \(O(BT^{2}d)\)          & \(O(log(BT^{2})shots)\) \\
Computation cost  & 1 fused-FLOP            & \(s\!\cdot\!\bigl(g\,t_{1\mathrm Q}+2t_{\mathrm{CX}}\bigr)+t_{\mathrm{ro}}\) \\
Runtime & $\,\lesssim 0.2\,\text{ms}$ & $\sim 0.1\,\text{s}$\\
Parallel scale         & SIMD / GPU cores        & feature per circuit \\
\bottomrule
\end{tabular}
\end{table}


\subsection{Model Evaluation}
\label{sec: me}
\begin{table}[t]
\centering
\begin{minipage}{\columnwidth}
\caption{Performance comparison of models on the Brown corpus. Note that each column is averaged over ten runs after 50 epochs. More qubits indicates larger hidden quantum dimensions as described in Eq.~\eqref{eq:qdim}.}
\centering
\begin{tabular}{c|cc|c|c}
\toprule
Model & QPL & Loss & Param & Qubit \\
\midrule
NanoGPT$^\dag$\citeyear{karpathy2023nanogpt} & 92.5 & 0.94 & 125K & -- \\
Q-LSTM \citeyear{chen2022quantum} & 125.3 & 1.18 & $\star$ & 6 \\
Quixer \citeyear{khatri2024quixer} & 117.1 & 1.61 & $\star$ & $\bullet$ \\
Hybrid QT \citeyear{smaldone2025hybrid} & 127.6 & 1.08 & $\star$ & $\bullet$ \\
\midrule
\textbf{VQT (default)} & \textbf{105.4} & \textbf{1.12} & $\star$ & \textbf{$\bullet$} \\
\textbf{VQT (large encoder)} & \textbf{108.2} & \textbf{1.08} & $\star$ & \textbf{12} \\
\bottomrule
\end{tabular}

\vspace{0.5em}
\begin{flushleft}
\footnotesize
$^\dag$ Benchmark baseline for QT model evaluation. \\
$\star$ Quantum models inherit classical layers and embedding. \\
$\bullet$ Default qubit size of 6 was selected for noisy QPU compatibility.
\end{flushleft}
\label{tab:performance_brown}
\end{minipage}
\end{table}

The Brown Corpus dataset \cite{ide2001american} comprises 1.01 million words across 500 text samples spanning 15 genres of American English prose. 
We utilize Fasttext \cite{bojanowski2016enriching} to compress the inputs as batch-processed matrices specified with three-dimensional tensors decoded for the quantum hidden dimension $Q_{dim}$. We refer to block A in Figure \ref{fig:circ_QC+EH} as the quantum oracle for the expressive quantum head. Note that, the default VQT quantum hidden dimension is 32 (see Eq. \eqref{eq:qdim}) with the qubits settings in Table \ref{tab:performance_brown} (we refer hyperparameter settings table for the VQT model evaluation in Appendix \ref{app:model} for more details).

The evidence demonstrates that VQT provides competitive results as a classical benchmark model denoted by NanoGPT (the smallest transformer). Note that we select the smallest GPT because of the limitation of current noisy qubits in the NISQ era. Interestingly, we observe that increasing the $Q_{dim}$ does not help with the perplexity of the model but slightly improves accuracy. However, we configure no dropout rate for VQT, as indicated in Appendix \ref{app:model}, which results in a lower loss rate than prior quantum models. This is because \vnqe\ allows nonlinear latent space transformation between classical and quantum layers, which enables the model to have a lower rate of overfitting problems, as discussed in \cite{kobayashi2022overfitting}.


\section{Discussion}
In conclusion, this study proposed a vectorized quantum model that allows data-encoded heuristic self-attention paradigm providing a end-to-end support for multi-head expansion in the future production quantum hardware. Unlike the prior quantum transformer method using learnable VQCs with parameter shift rule to calculate the gradients, our approach accurately and straightforwardly simulated the attention matrix using shot-based quantum circuit. By leveraging quantum data encoding techniques, the model largely ameliorate the overfitting problems, which enables a pathway for potential better reasoning model. We envision the quantum tokenizer might benefit our model at the scale.





\section{Acknowledgments}
This research used resources of the National Energy Research Scientific Computing Center, a DOE Office of Science User Facility, and UMD QLab. This study was supported by the Office of Science of the U.S. Department of Energy under Contract No. DE-AC02-05CH11231 using NERSC awards DDR-ERCAP0034486 and DDR-ERCAP0034385. All figures and numerical results reported in this study are publicly available at (https://github.com/gzquse/paper\_AAAI2025-quantum) to facilitate its reproducibility.

\bigskip
\bibliography{main}

\appendix
\section{Proof of Variance of VQDP}

Denote by \(b_i \in \{0,1\}\) the ancilla bit of shot \(i\) and map it to
\(Z_i = +1 - 2\,b_i\). With \(p = \Pr(b_i = 0) = \tfrac{1}{2}(1 + x\,y)\), the expectation and variance of each \(Z_i\) are
\begin{align}
    &\mathbb{E}[Z_i] = (+1)(p) + (-1)(1 - p) = 2p - 1 = x\,y, \\
    &\operatorname{Var}[Z_i] = \mathbb{E}[Z_i^2] - \mathbb{E}[Z_i]^2 = 1 - (x\,y)^2,
\end{align}
because \(Z_i \in \{+1, -1\}\), so \(Z_i^2 = 1\).
Since the \(Z_i\) are independent, the variance of their empirical mean is
\begin{equation}
  \operatorname{Var}[\hat Z]
  \;=\; \operatorname{Var}\!\left[\frac{1}{M} \sum_{i=1}^{M} Z_i \right]
  \;=\; \frac{1}{M^{2}} \sum_{i=1}^{M} \operatorname{Var}[Z_i]
  \;=\; \frac{1 - (x\,y)^2}{M}.
  \label{eq:variance}
\end{equation}
The bound for the deviation probability of the estimator \(\hat Z = \frac{1}{M} \sum Z_i\) can be derived more tightly from the binomial Chernoff inequality by applying it to the sum \(\sum b_i\), which is a binomial random variable with parameters \(M\) and \(q = \Pr(b_i = 1) = \frac{1}{2}(1 - x\,y)\). For any \(0 < \delta < 1\), the binomial Chernoff inequality provides
\begin{equation}
  \Pr\left[ \left| \frac{1}{M} \sum_{i=1}^{M} b_i - q \right| \geq \delta q \right]
  \;\leq\; 2 \exp\left( -\frac{\delta^2 q M}{3} \right).
  \label{eq:chernoff_b}
\end{equation}

Note that \(\hat Z = \frac{1}{M} \sum (1 - 2b_i) = 1 - 2 \cdot \frac{1}{M} \sum b_i\), so the deviation in \(\hat Z\) from its mean \(x\,y = 1 - 2q\) is directly tied to the deviation in the average of the \(b_i\) from \(q\). Therefore, by change of variables, for any \(\varepsilon > 0\),
\begin{align}
  \Pr\left[ |\hat Z - x\,y| \geq \varepsilon \right]
  &= \Pr\left[ \left| \frac{1}{M} \sum b_i - q \right| 
     \geq \frac{\varepsilon}{2} \right] \notag \\
  &\leq 2 \exp\left( -\frac{\varepsilon^2 M}{12 q} \right),
  \label{eq:chernoff_Z}
\end{align}
where the last step uses \(\delta = \varepsilon/(2q)\) and substitutes it into \eqref{eq:chernoff_b}, providing an exponentially decaying bound on the probability that the empirical estimator \(\hat Z\) deviates from its expectation \(x\,y\) by more than \(\varepsilon\), in terms of the number of shots \(M\).


\section{Model Settings}
\label{app:model}
\begin{table}[H]
\caption{Main hyperparameters for the vectorized quantum transformer (VQT) model configuration and training setup.}
\centering
{\small
\begin{tabular}{l|cc}
\toprule
\textbf{Hyperparameter} & \textbf{Default} & \textbf{Large}\\
\midrule
Vocabulary size ($V$) & 100 & 100 \\
Sequence length ($T$) & 6 & 8\\ 
Embedding dimension ($d$) & 32 & 32 \\
Feed-forward hidden size ($d_{\mathrm{ff}}$) & 128 & 128 \\
VQT blocks ($B$) & 1 & 1 \\
Number of QSE ($H$) & 2 & 2 \\
Address, data qubits ($nq_{\mathrm{addr}}, nq_{\mathrm{data}}$) & 3, 3 & 6, 6 \\
Encoded dimension $Q_{\mathrm{dim}}$ & $24$ & $384$ \\
Shots per $nq_{addr}$ ($S$) & 1024 & 1024 \\
AngleMLP hidden size ($d_{\mathrm{mlp}}$) & $128$ & $128$ \\
Dropout $(\rho)$ & 0 & 0\\
Optimizer & AdamW & AdamW \\
Learning rate ($\eta$) & $1 \times 10^{-3}$ & $1 \times 10^{-3}$ \\
Batch size ($B_{\mathrm{train}}$) & 5 & 5 \\
\bottomrule
\end{tabular}\\
}
\label{tab:hyperparameters_vqt}
\end{table}


\end{document}